# Patterns of Selection of Human Movements III: Energy Efficiency, Mechanical Advantage, and Walking Gait

**Stuart Hagler**
Oregon Health & Science University
Portland, OR, USA
haglers@ohsu.edu

**Abstract:** Human movements are physical processes combining the classical mechanics of the human body moving in space and the biomechanics of the muscles generating the forces acting on the body under sophisticated sensory-motor control. One way to characterize movement performance is through measures of energy efficiency that relate the mechanical energy of the body and metabolic energy expended by the muscles. We expect the practical utility of such measures to be greater when human subjects execute movements that maximize energy efficiency. We therefore seek to understand if and when subjects select movements with that maximizing energy efficiency. We proceed using a model-based approach to describe movements which perform a task requiring the body to add or remove external mechanical work to or from an object. We use the specific example of walking gaits doing external mechanical work by pulling a cart and estimate the relationship between the avg. walking speed and avg. step length. In the limit where no external work is done, we find that the estimated maximum energy efficiency walking gait is much slower than the walking gaits healthy adults typically select. We then modify the situation of the walking gait by introducing an idealized mechanical device that creates an adjustable mechanical advantage. The walking gaits that maximize the energy efficiency using the optimal mechanical advantage are again much slower than the walking gaits healthy adults typically select. We finally modify the situation so that the avg. walking speed is fixed and derive the pattern of the avg. step length and mechanical advantage that maximize energy efficiency.

1 Introduction

Human movements are physical processes combining the classical mechanics of the human body moving through space with the biomechanics of the muscles generating the required forces under the control of sophisticated, coordinated sensory-motor and cognitive processes. [1, 2] Due to the sophisticated nature of human movements, it can be difficult to characterize the quality of their performance, though concepts such as "efficiency," "economy," and "effectiveness" have been proposed. [3] One approach to characterizing performance is to ignore the sensory-motor and cognitive aspects of motor control, and focus on the physical mechanics of movement by using one or more measures of energy efficiency that relate the mechanical energy of the body moving through space with the metabolic energy of the muscles generating the forces. We expect the practical utility of such measures to be greater when human subjects are trying to execute movements with maximum energy efficiency. We therefore seek to understand if and when subjects select movements with the goal of maximizing energy efficiency by understanding the patterns of maximum energy efficiency movements and comparing them to observed human movement patterns. The goal of this project is to examine how maximizing energy efficiency shapes the patterns of human movements. The general idea is that we can use models of external mechanical work and metabolic energy expenditure to estimate human movement patterns that maximize measures of energy efficiency.

Energy minimization appears to play a role in movement selection, [4, 5] mechanisms for energy minimization during movement have been proposed, [5-7] and, in particular walking gait has been looked at as a minimization process. [5, 8-10] A closely related concept to that of energy minimization is maximum energy efficiency, and several measures of the energy efficiency of a movement have been proposed. [3, 11, 12] Measures of energy efficiency take their clearest form for movements which perform a task requiring the body to add or remove external mechanical work to or from an object. The subject must generate a fixed amount of external mechanical work determined by the task while expending some amount of metabolic energy determined by how the movement is executed. We can use measures of energy efficiency to characterize the relationship between the external mechanical work done and the metabolic energy expended. Naïvely, we might suppose that a subject would naturally select those movement patterns that minimize metabolic energy expended and maximize the energy efficiency. However we can imagine an extremely fatigued subject asked to carry out the same task as a refreshed subject. We expect the fatigued subject to have more motivation to carry out the subject with minimal expenditure of metabolic energy than the refreshed subject, and experience suggests the fatigued subject would carry out the task more slowly. If this argument holds, then refreshed subjects do not generally select movements that maximize energy efficiency but rather execute movements to maximize some measure of movement utility along the lines of [5] involving other goals not involving energy expenditure, and with the result that they complete the task in less time.

In this paper, we look at the patterns of movements that maximize energy efficiency. The approach we take is to use models of the metabolic energy of movements doing external mechanical work to construct models of the energy efficiency from which we derive the movement patterns that maximize the energy efficiency. We carry out this project out in four parts. In the first part (Sec. 2), we provide an outline of the formalism we use to describe the metabolic energies and external mechanical work, reviewing the relevant models developed in [5, 13] and defining four measures of energy efficiency following Gaesser & Brooks: [12] gross energy, net energy, work, and delta work efficiency. In the second part (Sec. 3), we review the model for walking gaits doing external mechanical work developed in [13]. We review the results in [5, 13] for how well the model accounts for the empirical data reported by Atzler & Herbst, [14] and argue for a modification to the fitting procedure that provides a good fit to the Atzler & Herbst data while still providing a good account of the relationship between the avg. walking speed and avg. step length of walking gaits observed by Grieve [15] in the case where no external work is done. In the third part (Sec. 4), we use the model to estimate the four measures of energy efficiency as functions of the external force for walking gaits doing external mechanical work. We find that the estimated work and delta work efficiencies agree with estimates in Donovan & Brooks. [16] We then derive the walking gaits that maximize the net energy efficiency, and find that in the limit as the external force goes to zero, these walking gaits become very slow and near to those observed for pre-rehabilitation Parkinson's disease (PD) subjects reported by Frazzitta et al. [17] In the fourth part (Sec. 5), we modify the situation in Sec. 4 by introducing an idealized mechanical device that creates an adjustable mechanical advantage. We again find that the walking gaits that PD subjects reported by Frazzitta et al. [17] We further look at the avg. step lengths and mechanical advantages that maximize the net energy efficiency given a fixed avg. walking speed.

## 2 Energy Efficiencies of a Movement

We begin with the first part of this project in which we develop models of the metabolic energy and external mechanical work of a movement, and relate them using four measures of energy efficiency. We begin by reviewing the approach to modeling human movements that we developed in [5, 13] including external mechanical work as developed in [13]. We then define four measures of energy efficiency of human movements following Gaesser & Brooks [12] that allow us to relate the metabolic energy of a movement to the external mechanical work done by the movement. We finally look at the measures of energy efficiency that we have developed as goals in movement selection.

*2.1 Metabolic Energy*

We model the human body executing a movement as in [5, 13] using a skeleton consisting of a system of N ("nu") segments attached at $N$ joints. A movement causes the joints to move with joint-trajectories so that the *n-th* joint moves with joint-trajectory $\vec{x}_n(t)$. The metabolic rate $\dot{W}(t)$ of expending metabolic energy is the sum of metabolic rates $\dot{W}_n(t)$ associated with each joint $\dot{W}(t) \approx \sum_{n=1}^{N} \dot{W}_n(t)$. The metabolic rate associated with a joint is given by a function:

$$\dot{W}_n(t) = \dot{W}_n^F\left(\vec{F}_n(t)\right) + \dot{W}_n^E\left(\vec{F}_n(t), \vec{v}_n(t)\right). \tag{1}$$

The metabolic rates $\dot{W}_n^F(t)$ and $\dot{W}_n^E(t)$ have the mathematical forms:

$$\begin{aligned}\dot{W}_n^F\left(\vec{F}_n(t)\right) &\approx \varepsilon_n F_n(t)^2, \\ \dot{W}_n^E\left(\vec{F}_n(t), \vec{v}_n(t)\right) &\approx \eta_n \vec{F}_n(t) \cdot \vec{v}_n(t).\end{aligned} \tag{2}$$

The quantities $\varepsilon_n$ and $\eta_n$ are constant parameter values characterizing the associated metabolic rates. The parameters $\eta_n$ may take on different values when the muscles add or remove mechanical energy to or from a segment, though we require they be constant in each case. The mechanical energy is *positive* when mechanical energy is added to a segment and *negative* when it is removed.

*2.2 External Mechanical Work*

The *external mechanical work* $U_{ext}(t)$ is mechanical energy transferred from the segments of the body to objects external to the body with the intent of carrying out a task using those objects. It should be distinguished from any sort of energy loss that is part of a movement but does directly contribute a specified task. For example, when a subject walks so as to push an object from one position to another where there is friction between the object and the floor, the mechanical energy the subject provides to the object to make it move is external mechanical work while mechanical energy lost at heel-strike during walking is not. This distinction is discussed further in [13]. The external mechanical work is *positive* when mechanical energy is added to an object and *negative* when it is removed.

*2.3 Measures of Energy Efficiency*

We define four measures of the energy efficiency following Gaesser & Brooks: [12] (i) *gross energy efficiency* $\upsilon_{gross}$ (external mechanical work done divided by gross metabolic energy expended), (ii) *net energy efficiency* $\upsilon_{net}$ (external mechanical work done divided by metabolic energy expended), (iii) *work efficiency* $\upsilon_{work}$ (external mechanical work done divided by metabolic energy expended above the case when no external mechanical work is done) , (iv) *delta work efficiency* $\upsilon_{delta}$ (change in external mechanical work done divided by change metabolic energy as external mechanical work rate changes). For a subject having a resting metabolic rate $\dot{W}_{rest}$ and a movement doing external mechanical work $U_{ext}$ in a time $T$, and expending metabolic energy of $W(U_{ext})$, we express the four measures of energy efficiency formally as:

$$\begin{aligned}
\upsilon_{gross}\left(U_{ext}\right) &= U_{ext} / \left(\dot{W}_{rest}T + W\left(U_{ext}\right)\right), \\
\upsilon_{net}\left(U_{ext}\right) &= U_{ext} / W\left(U_{ext}\right), \\
\upsilon_{work}\left(U_{ext}\right) &= U_{ext} / \left(W\left(U_{ext}\right) - W\left(0\right)\right), \\
\upsilon_{delta}\left(U_{ext}\right) &= \left(\frac{dW}{dU_{ext}}\left(U_{ext}\right)\right)^{-1}.
\end{aligned} \quad (3)$$

*2.4 Energy Efficiency and Movement Selection*

We are interested in when and if a subject selects a movement that maximizes any of the measures of energy efficiency in (3). For the subject to select a movement to maximize the gross energy efficiency, we must suppose that the subject is able to effectively estimate the resting or basal metabolic energy $\dot{W}_{rest}T$ relative to the metabolic energy $W(U_{ext})$. We are not aware of any mechanism for doing this, however, in [5] we did propose a mechanism using Stevens' power law [18, 19] by which a subject would be able to estimate the metabolic energy associated with generating muscle forces. We find below that the subject can estimate the net energy efficiency for walking gain described by the models in [5, 13] using only estimates of the perceived muscle forces, and therefore may be able to select walking gaits that maximize the net energy efficiency. Similarly, using estimates of the perceived muscle forces, the subject may be able to select movements that maximize the work efficiency. The form of the delta work efficiency is complicated and does not seem to apply to typical movements; however we find below that the work and delta work efficiencies are correlated in walking gait.

**3 Walking Gait**

In the second part of the project, we review the model walking gaits that we have developed in [5, 13]. We proceed first to review the development of the metabolic energy model of walking gait in [5, 13]: (i) we review the kinematic model of walking gait that we have previously used in [5, 13] (Sec. 3.2), (ii) we review and summarize the uncorrected metabolic energy model that describes one stop during walking gait doing external mechanical work developed in [13] (Sec. 3.3), (iii) we review the parameter estimates for the uncorrected metabolic energy model made in [5, 13] using available empirical data in Atzler & Herbst [14] for the cases of walking not doing external mechanical work [5] and doing external mechanical work, [13] respectively (Sec. 3.4). We then observe that the success of the uncorrected metabolic energy model using the parameter values estimated in [5] in accounting for empirical observations in Grieve [15] for the relationship between the avg. walking speed and avg. step length leads us to prefer the parameter values estimated in [5] over those estimated in [13] for the description of walking gait when no external mechanical work is done. We therefore: (iv) construct the form of a corrected metabolic energy model for walking gait doing external work which provides a formal account of the differences in the parameter values estimated in [5, 13] (Sec. 3.5), and (v) use this to fit the uncorrected metabolic energy model to the Atzler & Herbst data so that it has the parameter values in [5] when no external work is done and provides an approximate model when external work is done (Sec. 3.6).

*3.1 Some Anthropometric Values*

For convenience, we give here, in one place, a number of relevant anthropometric values. A subject with mass $M$ and height $H$ has a mass in each leg (i.e. thigh, shank, and foot) of about $\mu = rM$, and the length of the leg of about $L = \rho H$ where $r = 0.16$ and $\rho = 0.53$. [20] The mass of the torso carried by the stance leg during a step is $m = (1 - 2r)M$. The avg. walking speeds and avg. step lengths for adults aged 20-49 years are $v° \approx 1.3 \text{ m} \cdot \text{s}^{-1}$ and $s° \approx 0.61 \text{ m}$. [21]

*3.2 Kinematic Model*

We use the kinematic model for walking gaits developed in [5, 13]. This is a two segment model with one segment for each leg. The mass of the torso is located at the *torso* which is the point where the two segments meet; the mass of each leg is located in the feet at the far end of the leg segments from the torso. The torso maintains a constant height throughout the walking gait and maintains a constant speed along a straight line parallel to the ground. During one step, one leg is the *stance leg* which supports the torso as the torso moves over it, and the other is the *swing leg* which swings under the torso. The stance foot remains fixed on the ground while the swing foot glides a negligible distance above the ground; the legs lengthen or shorten as needed.

We only look at steady state walking gaits that are in progress and maintain constant values for the gait parameters; we do not look at the process of starting or stopping a walking gait. We describe walking gait using two gait parameters: (i) the avg. walking speed (the constant speed of the torso) $v$, and (ii) the avg. step length (the distance between the feet when they are both on the ground) $s$. We define the unit vector $\hat{v}$ to be the direction of motion of the torso, and use the model developed in [5] to describe the motions of the torso and swing foot:

$$\dot{\vec{x}}_{torso}(t) = v\hat{v},$$
$$\ddot{\vec{x}}_{foot}(t) \approx \begin{cases} \left(8v^2 \ / \ s\right)\hat{v}, & 0 \leq t \leq s \ / \ 2v, \\ -\left(8v^2 \ / \ s\right)\hat{v}, & s \ / \ 2v < t \leq s \ / \ v. \end{cases} \quad (4)$$

*3.3 Uncorrected Metabolic Energy Model*

We denote the muscle force applied by the stance leg to the torso by $\vec{F}'_{st}(t)$ and the force applied to the swing leg by $\vec{F}_{sw}(t)$. The torso is made to exert an external force $F_{ext}$ to perform external mechanical work. We define the external force $F_{ext}$ to oppose the horizontal motion of the torso so that it can be written $-F_{ext}\hat{v}$. The force the body must apply to compensate for the external force is $F_{ext}\hat{v}$. We find that $\vec{F}'_{st}(t) = \vec{F}_{st}(t) + F_{ext}\hat{v}$ where $\vec{F}_{st}(t)$ is the force of the stance leg on the torso when there is no external force. We associate parameters $\varepsilon_{st}$ and $\eta_{st}$ with the stance leg and a parameter $\varepsilon_{sw}$ with the swing leg; these parameters correspond to the parameters in (2). The time required to execute a step is $T = s/v$. The metabolic energy per step is the sum of a constant term $W_0$, and three metabolic energies: (i) the energy expended generating the force $\vec{F}'_{st}(t)$ of the stance leg, (ii) the energy expended generating the force $\vec{F}_{sw}(t)$ of the swing leg, and (iii) the energy expended by the stance leg to provide the mechanical energy of the external mechanical work; this is:

$$W\left(v, s, F_{ext}\right) \approx W_0 + \varepsilon_{st}\int_0^{s/v} F'_{st}(t)^2 \, dt + \varepsilon_{sw}\int_0^{s/v} F_{sw}(t)^2 \, dt + \eta_{st}F_{ext}v\int_0^{s/v} dt. \quad (5)$$

Expanding $F'_{st}(t)^2 = F_{st}(t)^2 + 2F_{ext}\vec{F}_{st}(t) \cdot \hat{v} + F_{ext}^2$ in (5), we find:

$$W(v,s,F_{ext}) \approx W_0 + \varepsilon_{st}\int_0^{s/v} F_{st}(t)^2 dt + \varepsilon_{sw}\int_0^{s/v} F_{sw}(t)^2 dt \\ + 2\varepsilon_{st}F_{ext}\int_0^{s/v} \vec{F}_{st}(t) \cdot \hat{v} dt + \left(\varepsilon_{st}F_{ext}^2 + \eta_{st}F_{ext}v\right)\int_0^{s/v} dt. \qquad (6)$$

The values of the parameters $W_0$, $\varepsilon_{st}$, $\varepsilon_{sw}$, and $\eta_{st}$ must be estimated empirically.

As we have shown in [13], for the movement in (4), the motion of the leg satisfies the relationships:

$$\begin{aligned} L\sin\theta(t) &= vt - s/2, \\ \vec{F}_{st}(t) &= (1/2)mg\sin 2\theta(t)\hat{v}, \\ \vec{F}'_{st}(t) &= (1/2)mg\sin 2\theta(t)\hat{v} + F_{ext}\hat{v}. \end{aligned} \qquad (7)$$

And, as we have shown in [5, 13], for the movement in (4), during the first half of the step, the motion of the swing leg satisfies the relationships:

$$\begin{aligned} L\sin\varphi(t) &= (4v^2/s)t^2 - vt - s/2, \quad 0 \leq t \leq s/2v, \\ \vec{F}_{sw}(t) &= \left((4\mu v^2/s) + (\mu g/2)\sin 2\varphi(t)\right)\hat{v}. \end{aligned} \qquad (8)$$

The motion of the swing leg is symmetric so the motion during the first half of the step suffices to allow us to make the calculations we need. Combining (6), (7), and (8), we find that, in the uncorrected metabolic energy model, the metabolic energy per step satisfies:

$$W(v,s,F_{ext}) \approx W_0 + \left(\varepsilon_{st}m^2g^2/12L^2 + \varepsilon_{sw}\mu^2g^2/5L^2\right)s^3/v \\ - \left(10\varepsilon_{sw}\mu^2g/3L\right)vs + \left(16\varepsilon_{sw}\mu^2\right)v^3/s \\ + \left(\varepsilon_{st}\right)F_{ext}^2s/v + \left(\eta_{st}\right)F_{ext}s. \qquad (9)$$

We find it adds to conceptual clarity to define the two quantities:

$$\begin{aligned} W_{gait}(v,s) &= W(v,s,0), \\ W_{ext}(v,s,F_{ext}) &= W(v,s,F_{ext}) - W(v,s,0). \end{aligned} \qquad (10)$$

In this way, $W_{gait}(v,s)$ gives the metabolic energy per step simply to carry out the walking gait while $W_{ext}(v,s,F_{ext})$ gives the metabolic energy per step beyond $W_{gait}(v,s)$ needed to generate the external force $F_{ext}$. We can use (10) to write (9) compactly as:

$$W(v,s,F_{ext}) \approx W_{gait}(v,s) + W_{ext}(v,s,F_{ext}). \qquad (11)$$

*3.4 Review of Empirical Studies (Atzler & Herbst, 1927)*

We now review the results of [5, 13] for making the uncorrected metabolic energy model in (9) into an estimator of the metabolic energy during walking over the range of allowed walking gaits by using empirical data to produce estimates for the values of the parameters $W_0$, $\varepsilon_{st}$, $\varepsilon_{sw}$, and $\eta_{st}$.

Atzler & Herbst [14, 22] observed one subject (male, $M = 68$ kg, $H = 1.7$ m, aged 39 years, mass-normalized resting metabolic rate $\dot{w}_{rest} = \dot{W}_{rest}/M = 0.30$ cal·kg$^{-1}$·s$^{-1}$) perform a variety walking gaits, and measured the metabolic energy for each walking gait. The subject was trained to walk on a horizontal treadmill using all 20 combinations of 4 avg. step lengths $s$ (0.46 m, 0.60 m, 0.76 m, and 0.90 m) and 5 avg. cadences $v/s$ (0.83 step·s$^{-1}$, 1.25 step·s$^{-1}$, 1.7 step·s$^{-1}$, 2.2 step·s$^{-1}$, and 2.5 step·s$^{-1}$). In addition to walking freely, the subject walked pulling a cart (Deichselwagen) such that the subject had to apply four different external forces $F_{ext}$ to overcome friction (100 N, 110 N, 130 N, and 160 N). The handle by which the subject pulled the cart attached to the cart via a rigid shaft that was fixed so that the handle was positioned 1.0 m above the floor. Some combinations of walking speed, cadence, and external force were left out of the study. For the setups requiring external force $F_{ext} = 130$ N, the subject performed 18 of the 20 combinations, leaving out walking gaits with avg. step lengths and avg. cadences of (i) $s = 0.90$ m and $v/s = 2.2$ step/s, and (ii) $s = 0.90$ m and $v/s = 2.5$ step/s. For the setups requiring external force $F_{ext} = 160$ N, the subject performed 12 of the 20 combinations, leaving out walking gaits with avg. step length of 0.90 m or avg. cadence 0.83 step·s$^{-1}$.

Rewriting the uncorrected metabolic energy model in (9), we find:

$$W \approx W_0 + \left(m^2 g^2 s^3 / 12L^2 v + F_{ext}^2 s / v\right)\varepsilon_{st} \\ + \left(16\mu^2 v^3 / s - 10\mu^2 gvs / 3L + \mu^2 g^2 s^3 / 5L^2 v\right)\varepsilon_{sw} + \left(F_{ext} s\right)\eta_{st}. \quad (12)$$

In [5], we fit the uncorrected metabolic energy model using only the data for the case where $F_{ext} = 0$. Using ordinary least-squares (OLS) regression, this model fit the data for Atzler & Herbst's subject with $R^2 = 0.99$ and $p < 0.0001$, and the parameter values were:

$$\begin{aligned} W_0 &\approx 9.0\,cal, \\ \varepsilon_{st} &\approx 2.5 \times 10^{-3}\,cal \cdot N^{-2} \cdot s^{-1}, \\ \varepsilon_{sw} &\approx 1.7 \times 10^{-3}\,cal \cdot N^{-2} \cdot s^{-1}. \end{aligned} \quad (13)$$

Inspection of the 95% confidence intervals showed that all 3 parameters were statistically significant. In [13], we fit the uncorrected metabolic energy model using all the data. Using OLS regression, this model fit the data for Atzler & Herbst's subject with $R^2 = 0.96$ and $p < 0.0001$, and the parameter values were:

$$\begin{aligned} W_0 &\approx 7.0\,cal, \\ \varepsilon_{st} &\approx 1.9 \times 10^{-3}\,cal \cdot N^{-2} \cdot s^{-1}, \\ \varepsilon_{sw} &\approx 2.6 \times 10^{-3}\,cal \cdot N^{-2} \cdot s^{-1}, \\ \eta_{st} &\approx 0.62\,cal \cdot J^{-1}. \end{aligned} \quad (14)$$

Inspection of the 95% confidence intervals showed that all 4 parameters were statistically significant.

*3.5 Corrected Metabolic Energy Model*

In [5], we used the parameter values in (13) to produce a maximum utility model of walking gaits that was able to produce a relationship between the avg. walking speed and avg. step length that approximated the relationship observed in Grieve. [15] We find however that the corresponding parameter values in (14) when used to produce a model to approximate that relationship perform less well in approximating the relationship observed in Grieve. We argue that the model overfits to the cases where there is an external force and therefore has a poorer fit for the cases where no external mechanical work is done. We further argue that this is due to the uncorrected metabolic energy model providing an incomplete description of how the presence of an external force affects the metabolic energy of walking gaits. We can complete the description formally by introducing an unknown function $W_{ext}^{(1)}(v, s, F_{ext})$ that is required to satisfy $W_{ext}^{(1)}(v, s, 0) = 0$. We therefore expect the uncorrected metabolic energy model in (11) to be replaced with a corrected metabolic energy model of the form:

$$W(v, s, F_{ext}) = W_{gait}(v, s) + W_{ext}(v, s, F_{ext}) \\ + W_{ext}^{(1)}(v, s, F_{ext}). \quad (15)$$

The $W_{gait}(v, s)$ term in (15) should provide a correct description of walking gait when there is no external force $F_{ext}$ and therefore is better fit using the parameters in (13) rather than those in (14). However for the cases in which there is an external force $F_{ext}$ around the size of those observed by Atzler & Herbst, we expect the uncorrected metabolic energy using the parameters (14) to provide a better approximation of the corrected metabolic energy model in (15). The uncorrected metabolic energy model provided good fits in both the case where the external force $F_{ext} = 0$ and the cases in which there was an external force $F_{ext}$ with parameters in (13) and (14) that take on similar values. We therefore expect the contribution of $W_{ext}^{(1)}(v, s, F_{ext})$ to the corrected metabolic energy model to be small.

*3.6 Empirical Study (Atzler & Herbst, 1927)*

We would like to produce a model that reduces to the uncorrected metabolic energy model with parameter values in (13) when there is no external force $F_{ext}$, but also provides estimates when there is an external force. To do this we use the parameter values for $\varepsilon_{st}$ and $\varepsilon_{sw}$ in (13) and fit parameter values for $W_0$ and $\eta_{st}$ to the full Atzler & Herbst data set used in Sec. 3.4 using the model:

$$W - \left(m^2 g^2 s^3 / 12 L^2 v + F_{ext}^2 s / v\right) \varepsilon_{st} \\ - \left(16 \mu^2 v^3 / s - 10 \mu^2 g v s / 3L + \mu^2 g^2 s^3 / 5 L^2 v\right) \varepsilon_{sw} \approx W_0 + \left(F_{ext} s\right) \eta_{st}. \quad (16)$$

Using OLS regression, this model fit the data for Atzler & Herbst's subject with $R^2 = 0.81$ and $p < 0.0001$. Inspection of the 95% confidence intervals showed that both parameters were statistically significant. The resulting model parameters were:

$$\begin{aligned} W_0 &\approx 5.7 \, cal, \\ \varepsilon_{st} &\approx 2.5 \times 10^{-3} \, cal \cdot N^{-2} \cdot s^{-1}, \\ \varepsilon_{sw} &\approx 1.7 \times 10^{-3} \, cal \cdot N^{-2} \cdot s^{-1}, \\ \eta_{st} &\approx 0.53 \, cal \cdot J^{-1}. \end{aligned} \quad (17)$$

When we apply this model to the full data set available to Atzler & Herbst, we estimate that the model performs with $R^2 = 0.93$ and $p < 0.0001$. The fit is illustrated in Fig. 1 using all the walking gaits observed by Atzler & Herbst.

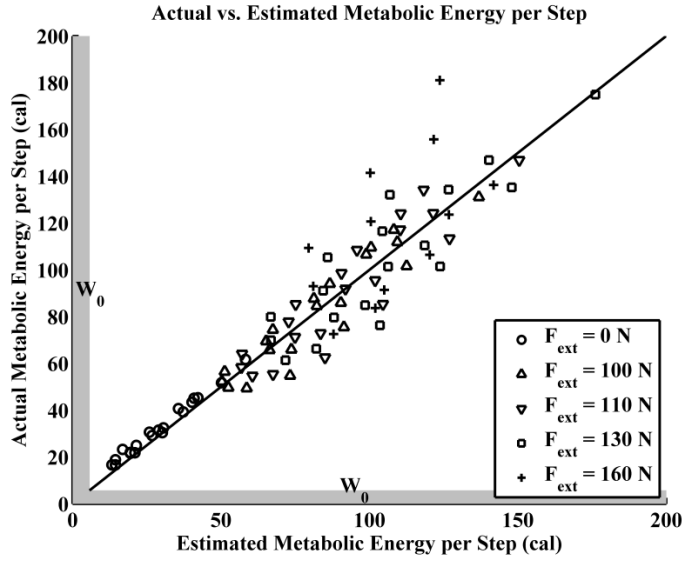

**Figure 1.** Actual vs. estimated metabolic energy per step. We use the uncorrected metabolic energy model in (11) with the parameter values in (17) to estimate the metabolic energy for the 90 walking gaits in Atzler & Herbst; the estimated fit for the model for these walking gaits has $R^2 = 0.93$ and $p < 0.0001$. The value for the constant parameter $W_0$ is indicated. For reference, we show a segment of the line with slope one passing through the origin.

We attribute the systematic deviations of the estimated metabolic energy per step from the actual metabolic energy per step in Fig. 1 to the absence of the function $W_{ext}^{(1)}(v, s, F_{ext})$ in the uncorrected metabolic energy model. We may obtain some suggestion of the functional form of $W_{ext}^{(1)}(v, s, F_{ext})$ by looking at these deviations as functions of the avg. walking speed $v$, the avg. step length $s$, and the external force $F_{ext}$; we do this in Fig. 2.

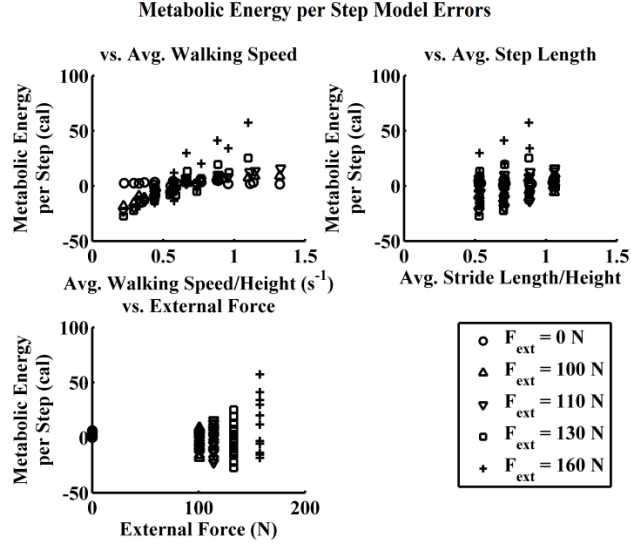

**Figure 2.** Metabolic energy per step model errors. We suggest the functional form of $W^{(1)}_{ext}(v, s, F_{ext})$ by plotting the differences in the actual metabolic energy per step and the metabolic energy per step estimated use the uncorrected metabolic energy model in (11) with the parameter values in (17) as functions of the avg. walking speed, the avg. step length, and the external force.

## 4 Energy Efficiency of Walking Gait

In the third part of this project, we use the metabolic energy and external mechanical work models to estimate the energy efficiency of walking gait. We first calculate the functional forms of the gross energy, net energy, work, and delta work efficiencies for walking gaits doing external mechanical work, and compare the resulting energy efficiency estimates to values reported by Donovan & Brooks. [16] We then estimate the relationship between the avg. walking speed and avg. step length for a subject selecting walking gaits doing external mechanical work of maximum net energy efficiency and find that they are determined by the external force $F_{ext}$. In the limit as the external force goes to zero, the walking gaits become very slow and near to those observed reported by Frazzitta et al. [17] for pre-rehabilitation PD subjects.

*4.1 Measures of Energy Efficiency*

We now calculate the gross energy, net energy, work, and delta work efficiencies of walking gaits doing external mechanical work using the definitions in (3). The external mechanical work per step is $F_{ext}s$ and the metabolic energy is $W(v, s, F_{ext})$, and the measures of energy efficiency for walking gaits doing external mechanical work are:

$$\begin{aligned}
\upsilon_{gross}(v, s, F_{ext}) &= F_{ext}s \,/\, \left(\left(\dot{W}_{rest}\right)s \,/\, v + W(v, s, F_{ext})\right), \\
\upsilon_{net}(v, s, F_{ext}) &= F_{ext}s \,/\, W(v, s, F_{ext}), \\
\upsilon_{work}(v, s, F_{ext}) &= \left(\eta_{st} + \left(\varepsilon_{st}\right)F_{ext} \,/\, v\right)^{-1}, \\
\upsilon_{delta}(v, s, F_{ext}) &= \left(\eta_{st} + \left(2\varepsilon_{st}\right)F_{ext} \,/\, v\right)^{-1}.
\end{aligned} \tag{18}$$

We illustrate the behavior of these measures by providing estimates for typical efficiencies using the uncorrected metabolic energy model in (9); we look at Atzler & Herbst's subject in Fig. 3. We provide two plots, one using the parameter values in (17), and a second using the parameter values in (14) to provide an estimate on the how much the first model deviates from the corrected metabolic energy model for external forces $F_{ext}$ on the order of those observed in Atzler & Herbst. Although we expect subjects to change the avg. walking speed $v$ and avg. step length $s$ of their walking gait as the external force $F_{ext}$ changes, we graph the energy efficiency measures assuming the subject walks using a walking gait with avg. walking speed $v°$ and avg. step length $s°$ for all external forces. The work and delta work efficiencies have close values and a similar functional form across the given range of external force; this supports the assertion that we made in Sec. 2.4 regarding these energy efficiencies.

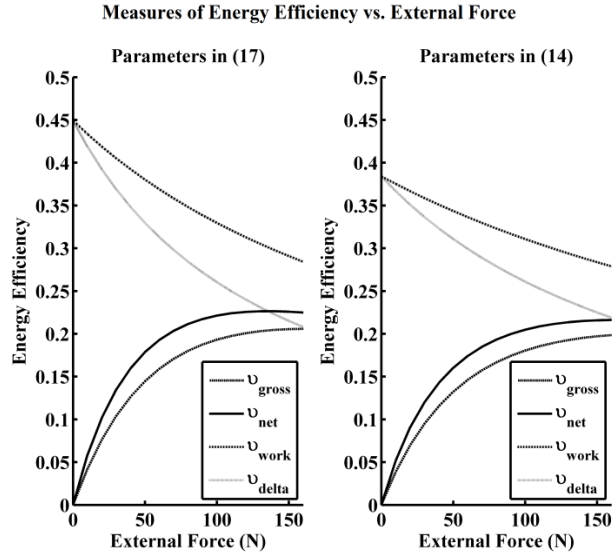

**Figure 3.** Measures of energy efficiency vs. external force. We give estimated functional forms of the gross, net, work, and delta energy efficiencies as the external force changes for a fixed walking gait with avg. walking speed $v°$ and avg. step length $s°$. We plot the uncorrected metabolic energy model with parameters in (17) as well as the uncorrected metabolic energy model with parameters in (14) to provide an estimate on the how much the first model deviates from the corrected metabolic energy model for external forces on the order of those observed in Atzler & Herbst.

We can compare the values estimated for the work, and delta work efficiencies in Fig. 3 to values measured empirically. Donovan & Brooks [16] observed subjects walking against vertical (gradient) and horizontal (trailing) force during steady-rate exercise, they estimated their subjects to have work efficiencies falling from about 0.32 to about 0.26 as the external force increased, and delta energy efficiencies falling from about 0.32 to 0.20 as the external force increased over the same range. These observations agree with those illustrated in Fig. 3 for each of the two parameter sets (17) and (14).

*4.2 Maximizing Net Energy Efficiency*

We look at the selection of a walking gait doing external mechanical work in which the subject's only goal is to maximize the net energy efficiency while carrying out the task. We suppose the subject must pull a cart requiring an external force $F_{ext}$ over a distance $D$. If the subject walks with avg. walking speed $v$ and avg. step length $s$, then the subject must take $N = D/s$ steps, and the total metabolic energy expended is $W_{total}(v, s, F_{ext}) = (D/s)W(v, s, F_{ext})$. We define the metabolic energy per unit distance $\Phi(v, s, F_{ext}) = W_{total}(v, s, F_{ext})/D$; this gives:

$$\Phi\left(v, s, F_{ext}\right) = W\left(v, s, F_{ext}\right)/s. \tag{19}$$

Looking at the definition of the net energy efficiency $\upsilon_{net}(v, s, F_{ext})$ in (18), we find that:

$$\upsilon_{net}\left(v, s, F_{ext}\right) = F_{ext} / \Phi\left(v, s, F_{ext}\right). \tag{20}$$

Therefore for fixed external force $F_{ext}$, minimizing $\Phi(v, s, F_{ext})$ maximizes the net energy efficiency $\upsilon_{net}(v, s, F_{ext})$, and we find that minimizing the total metabolic energy expended is equivalent to maximizing the net energy efficiency.

The subject selects the walking gait that minimizes $\Phi(v, s, F_{ext})$ given an external force $F_{ext}$; this minimum is the solution to the system of equations:

$$\begin{aligned}\frac{\partial \Phi}{\partial v}(v, s, F_{ext}) &= 0, \quad (A) \\ \frac{\partial \Phi}{\partial s}(v, s, F_{ext}) &= 0. \quad (B)\end{aligned} \tag{21}$$

We can write the uncorrected metabolic energy model in (9) divided by the avg. step length $s$ compactly as $\Phi(v, s, F_{ext}) \approx \eta_{ext} F_{ext} + \alpha s^2/v - \beta v + \gamma v^3/s^2 + \delta/s + \varepsilon F_{ext}^2/v$ where $\alpha$, $\beta$, $\gamma$, $\delta$, and $\varepsilon$ are constants for the subject. The $\varepsilon$ is used here for notational convenience should not be confused with the parameter defined in (2) or appearing in any of the models obtained from (2). We can now rewrite (21) as the system of equations:

$$\begin{aligned}\alpha s^4 + \beta v^2 s^2 - 3\gamma v^4 &= -\varepsilon F_{ext}^2 s^2, \quad (A) \\ 2\alpha s^4 - \delta v s - 2\gamma v^4 &= 0. \quad (B)\end{aligned} \tag{22}$$

The walking gait doing external mechanical work of maximum net energy efficiency is given by $v^* = v(F_{ext})$ and $s^* = s(F_{ext})$ solving the system of equations in (22). No term in (22) contains a factor of $\eta_{ext}$, therefore the subject does not need to estimate the metabolic energy expended generating mechanical energy in order to maximize the net energy efficiency; this supports the assertion that we made in Sec. 2.4 regarding the subject's possible ability to estimate the net energy efficiency using only estimates of the perceived muscle forces.

*4.3 Predicted Movement Patterns*

We illustrate $\Phi(v, s, F_{ext})$ for external forces $F_{ext}$ of 0, 100 N, 110 N, and 130 N in Fig. 3. We have used the avg. stride length rather than the avg. step length in Fig. 3 to allow for comparison to the corresponding figure in [5] as well as in Grieve. [15] The contour lines indicate constant $\Phi$ in the $\Phi$ landscape given by the uncorrected metabolic energy model in (9) using the parameter values in (17). The height-normalized typical adult walking gait is indicated with avg. walking speed $v°/H$ and avg. stride length $2s°/H$. The dark grey areas denote regions where, as we have argued in [5, 13], walking gaits do not occur but that any bipedal forms of locomotion occurring in these regions should be considered distinct from walking gaits. The white area denotes the region of walking gaits while the light grey region denotes the region of very slow walking gaits defined in [13]. The solid black lines indicate the limits defining the various regions again defined in [13].

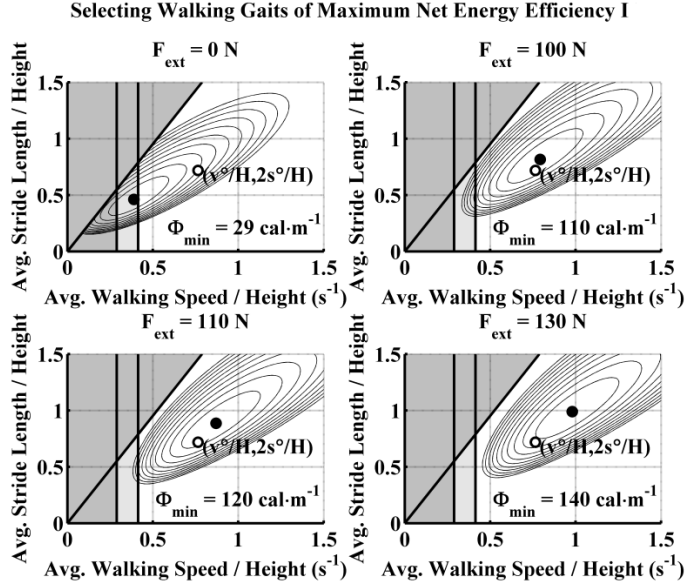

**Figure 4.** Selecting walking gaits of maximum net energy efficiency I. We plot the metabolic energy per step landscape for external forces $F_{ext} = 0$, 100 N, 110 N, and 130 N. The contour lines indicate constant metabolic energies per step in cal·m$^{-1}$ in the landscape. The black circle indicates the minimum metabolic energy per step. The contour lines appear at increments of 2 cal·m$^{-1}$ from that minimum. The typical adult walking gait is indicated with height-normalized avg. walking speed $v°/H$ and avg. stride length $2s°/H$. The dark grey areas denote regions where walking gaits do not occur. The white are denotes the region of walking gaits while the light grey region denotes the region of very slow walking gaits. The solid black lines indicate the limits defining the various regions.

We illustrate the curve B corresponding to the second equation in (22) in Fig. 4. We have used the height-normalized avg. stride length rather than the avg. step length in Fig. 4 to allow for comparison to the corresponding figure in [5] as well as in Grieve. [15] The values for the parameters $\alpha$, $\beta$, $\gamma$, $\delta$, and $\varepsilon$ are those for Atzler & Herbst's subject estimated in (17) and (14) as indicated. We plot two estimates of the B curve in (22), one using the uncorrected metabolic energy model with the parameter values in (17), and a second using the uncorrected metabolic energy model with parameter values in (14) to provide an estimate on the how much the first model deviates from the corrected metabolic energy model for external forces $F_{ext}$ on the order of those observed in Atzler & Herbst; the curves are labeled B(17) and B(14), respectively. The curves B(17) and B(14) are parameterized by the external force $F_{ext}$, and we have indicated the points at which $F_{ext} = 0$ and 160 N for both curves. We also give a curve indicating the relationship between the avg. walking speed and avg. stride length for an adult that exhibits the typical walking gait has been estimated for it using the model for selection of walking gaits developed in [5] with the parameter values in (17). The height-normalized typical adult walking gait is indicated with avg. walking speed $v°/H$ and avg. stride length $2s°/H$. As above, the dark grey areas denote regions where as we have argued in [5, 13], walking gaits do not occur but that any bipedal forms of locomotion occurring in these regions should be considered distinct from walking gaits. The white are denotes the region of walking gaits while the light grey region denotes the region of very slow walking gaits. The solid black lines indicate the limits defining the various regions.

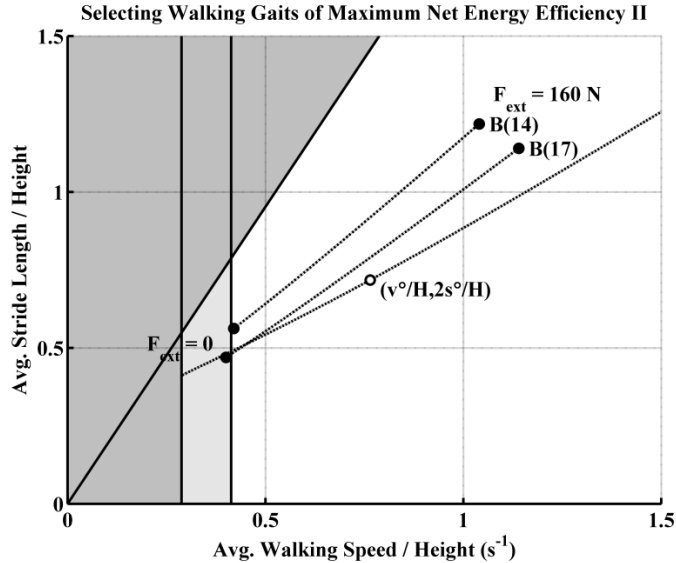

**Figure 5.** Selecting walking gaits of maximum net energy efficiency II. We plot two estimates of the B curve in (22), one using the simple metabolic energy model with the parameter values in (17), and with parameter values in (14); the curves are labeled B(17) and B(14), respectively. We have indicated the points at which $\boldsymbol{F_{ext}}$ = 0 and 160 N, respectively for both curves. The typical adult walking gait is indicated with height-normalized avg. walking speed $\boldsymbol{v°/H}$ and avg. stride length $\boldsymbol{2s°/H}$, and a curve estimating the relationship between the avg. walking speed and avg. stride length for adult subjects that exhibit the typical walking gait has been estimated for it using the parameter values in (17). The dark grey areas denote regions where walking gaits do not occur but that any bipedal forms of locomotion occurring in these regions should be considered distinct from walking gaits. The white are denotes the region of walking gaits while the light grey region denotes the region of very slow walking gaits. The solid black lines indicate the limits defining the various regions.

In the limit as the external force $F_{ext} \to 0$, the walking gait with maximum net energy efficiency is much slower than a typical adult walking gait. However this walking gait compare well to walking gaits observed for pre-rehabilitation PD subjects reported by Frazzitta et al. [17] These subjects were reported to have avg. walking speeds of $v \approx 0.50$ m·s$^{-1}$ to $0.60$ m·s$^{-1}$, or $v/H \approx 0.29$ step·s$^{-1}$ to $0.35$ step·s$^{-1}$ when height-normalized using an avg. height of 1.7 m, and a stride cycle of 0.6 step·s$^{-1}$, and thus avg. stride lengths of $2s \approx 0.83$ m to 1.0 m, or $2s/H \approx 0.49$ to $0.59$ when height-normalized using an avg. height of 1.7 m. This places the pre-habilitation walking gaits for these PD subjects close to the estimated walking gait with maximum net energy efficiency when no external mechanical work is done.

*4.4 Discussion*

In Fig. 5, for the lower external forces $F_{ext}$ the avg. walking speed $v$ that maximizes the net energy efficiency of the movement is slower than the typical avg. walking speed $v°$ while for the higher external forces have avg. walking speeds higher than $v°$. As this does not seem correspond to how we might intuitively expect a subject to select walking gaits doing external mechanical work, it does not appear that subjects typically select walking gaits doing external mechanical work that maximize the net energy efficiency, but rather execute movements to maximize some measure of movement utility along the lines of [5] involving other goals not involving energy expenditure, and with the result that they complete the task in less time.

The pre-rehabilitation PD subjects observed by Frazzitta et al. appear to have adopted walking gaits near those that minimize $W(v, s, 0)/s$. This suggests that subjects have a mechanism to self-select walking gaits that minimize this quantity at least in this limiting case. While it is possible that subjects can either directly or indirectly estimate $W(v, s, 0)/s$, an alternative mechanism would be for subjects to simply select very slow walking gaits in the light grey region in Fig. 5 by physically noting the behavior of the body at the limits of the region of very slow walking gaits described in [13]. At the upper limit of the light grey region, the subject would note when the torso seems to lose a large amount of mechanical energy with each stop, while at the lower limit of the light grey region (i.e. around the minimum avg. walking speed $v_{min}$ defined in [13]) the subject would note when the entire body seems to lose all of its mechanical energy with each step.

**5 Mechanical Advantage and Walking Gait**

In the fourth part of this project, we take the cart-pulling setup in Sec. 4, and introduce an idealized mechanical device that creates an adjustable mechanical advantage. We first calculate the avg. walking speed, avg. step length, and mechanical advantage of the system that give a walking gait doing external work with maximum net energy efficiency; we find that the selected walking gait and mechanical advantage have the subject walking with the lower limit avg. walking speed for very slow walking gaits and with the cart moving very slowly. We then modify the situation so that the subject walks with fixed avg. walking speed that is typical of adult walking gait and again calculate the avg. step length and mechanical advantage; in this case, the cart moves with a faster avg. speed.

*5.1 External Mechanical Work with Mechanical Advantage*

We imagine the subject is pulling the cart using an idealized mechanical device consisting of two spools of different radii that attach to each other so that they rotate at the same frequency around a common axle. Each spool is massless and each is attached to a length of massless cord of infinitesimal thickness which can be wound around the spool; the axle is frictionless. The subject walks with a steady state walking gait with avg. walking speed $v$ and avg. step length $s$ pulling on one of the cords; this cord is wound around a spool of radius $R$ and as the subject pulls the cord it unwinds from the spool causing both spools to rotate. The cart is attached to the other cord; as the spools rotate, this second cord winds itself around a spool of radius $r$ pulling the cart with an avg. speed $v_{cart}$; the cart requires a force $F_{cart}$ be applied to move at a constant avg. speed. The subject is free to adjust the mechanical advantage by swapping out the spool attached to the cart with spools of arbitrary radius $r$. As we did in the walking gait model developed in Sec. 3, we only look at steady state walking gaits that are in progress and maintain constant values for the gait parameters; we do not look at the process of starting or stopping a walking gait.

The cart requires a force $F_{cart}$ be applied to it for it to move at an avg. speed $v_{cart}$. The torque $\tau$ required on the spool of radius $r$ to pull the cart is $\tau = F_{cart} r$, so the subject must apply an external force satisfying $\tau = F_{ext} R$ to generate the required $F_{cart}$. In one rotation, the cord attached to the spool with radius $R$ travels a distance $2\pi R$ while that attached to the spool with radius $r$ travels a distance $2\pi r$. If one rotation takes a time $T$, the avg. speed of the torso is $v = 2\pi R/T$ and the avg. speed of the cart is $v_{cart} = 2\pi r/T$; we find:

$$v_{cart}\,/\,v = F_{ext}\,/\,F_{cart} = r\,/\,R. \tag{23}$$

The cost of adjusting the mechanical advantage $F_{ext}/F_{cart}$ so that the subject need only generate a lower external fore $F_{ext}$ is that the cart moves slower. In one rotation of the spools the cart receives a mechanical energy of $U_{cart} = 2\pi F_{cart} r$, while the subject generates a mechanical energy of $U_{ext} = 2\pi F_{ext} R$, so we find $U_{cart} = U_{ext}$. Thus all the generated mechanical energy above that which is a part of walking gait when no external mechanical work is done is given to the cart as external mechanical work. Combining (11) and (23) gives a metabolic energy per step of:

$$\begin{aligned}W(v,s,r,F_{cart}) \approx W_{gait}(v,s) + (\varepsilon_{st})(r\,/\,R)^2\,F_{cart}^2 s\,/\,v \\ + (\eta_{st})(r\,/\,R)F_{cart}s.\end{aligned} \tag{24}$$

*5.2 Maximizing Net Energy Efficiency*

We again suppose that the subject must move the cart a distance $D$. The subject must turn the spool attached to the cord attached to the cart $n = D/2\pi r$ times, and thus must move the torso a distance $d = 2\pi n R$, and therefore must take $N = d/s$ steps. Combining these gives $N = RD/rs$, and the total metabolic energy expended moving the cart the distance $D$ is $W_{total}(v,s,r,F_{ext}) = NW(v,s,r,F_{ext}) = (RD/rs)W(v,s,r,F_{ext})$. Again $\Phi(v,s,r,F_{ext}) = W_{total}(v,s,r,F_{ext})/D$; and we find:

$$\Phi(v,s,r,F_{cart}) = (R\,/\,r)(W(v,s,r,F_{cart})\,/\,s). \tag{25}$$

The walking gait doing external work expending the minimum metabolic energy to move the cart solves the system of equations:

$$\begin{aligned}
\frac{\partial \Phi}{\partial v}(v,s,r,F_{cart}) &= 0, \quad (A) \\
\frac{\partial \Phi}{\partial s}(v,s,r,F_{cart}) &= 0, \quad (B) \\
\frac{\partial \Phi}{\partial r}(v,s,r,F_{cart}) &= 0. \quad (C)
\end{aligned} \quad (26)$$

We can write the simple metabolic energy model in (25) divided by the avg. step length compactly in the form $\Phi(v,s,r,F_{cart}) \approx \eta_{ext} F_{cart} + \alpha \cdot (R/r)s^2/v - \beta \cdot (R/r)v + \gamma \cdot (R/r)v^3/s^2 + \delta \cdot (R/r)/s + \varepsilon \cdot (r/R)F_{cart}^2/v$ where $\alpha$, $\beta$, $\gamma$, $\delta$, and $\varepsilon$ are constants for the subject. The $\varepsilon$ is used here for notational convenience should not be confused with the parameter defined in (2) or appearing in any of the models obtained from (2). We can now rewrite (21) as the system of equations:

$$\begin{aligned}
\alpha s^4 + \beta v^2 s^2 - 3\gamma v^4 &= -\varepsilon \cdot (r/R)^2 F_{cart}^2 s^2, \quad (A) \\
2\alpha s^4 - \delta vs - 2\gamma v^4 &= 0, \quad (B) \\
\alpha s^4 - \beta v^2 s^2 + \delta vs + \gamma v^4 &= \varepsilon \cdot (r/R)^2 F_{cart}^2 s^2. \quad (C)
\end{aligned} \quad (27)$$

Equations B in (22) and (27) are the same so the walking gaits solving (26) lie along the curve B illustrated in Fig. 5. We note that no term in (27) contains a factor of $\eta_{ext}$, therefore the subject again does not need to estimate the metabolic energy expended generating mechanical energy for the external mechanical work in order to maximize the net energy efficiency.

The system of equations in (27) only has a solution at the origin. However, in practice, the model we have developed is no longer valid below a minimum avg. walking speed $v_{min}$ that we have derived in [13] and indicated in Figs. 4 and 5 as the lower limit the domain of very slow walking gaits. If we require the subject to use a walking gait, then the subject should walk with avg. walking speed $v_{min}$; we may find the avg. step length and mechanical advantage by fixing the avg. walking speed to $v_{min}$, and using the results developed in Sec. 5.3. This walking gait is again comparable to walking gaits observed for pre-rehabilitation PD subjects reported by Frazzitta et al. [17] We have argued in [13] that we should define *walking gaits* so that the torso never comes to a complete stop during the gait cycle, $v_{min}$ represents a lower limit on the avg. walking speed of such walking gaits. We have discussed this limit other forms of bipedal locomotion that exist at lower avg. walking speeds in [13]; such still slower gaits would be described by a different model from the one that we have developed in this paper.

*5.3 Maximizing Net Energy Efficiency Given a Fixed Avg. Walking Speed*

While, in the approach we have developed in [5, 13], it would be more correct to construct a maximum utility model that includes a goal that would directly or indirectly indicate that the subject desires to walk with a faster avg. walking speed, we adopt the approach of fixing the avg. walking speed to $v^*$ to keep the treatment straightforward. In this case, noting that the average metabolic power $\langle \dot{W}_{gait}(v,s) \rangle$ of walking gait with no external load is given by $\langle \dot{W}_{gait}(v,s) \rangle = W_{gait}(v,s) \cdot (v/s)$, the system of equations in (27) reduces to:

$$\begin{aligned}
2\alpha s^4 - \delta v^* s - 2\gamma \cdot (v^*)^4 &= 0, \quad (B) \\
r/R &= \sqrt{\langle \dot{W}_{gait}(v^*,s) \rangle / \varepsilon_{st}} / F_{cart}. \quad (C)
\end{aligned} \quad (28)$$

The walking gait of maximum net energy efficiency is given by $s^* = s(v^*)$ and $r^* = r(v^*, F_{cart})$ solving the system of equations in (28). The avg. step length $s^*$ is given by curve B illustrated in Fig. 5 at the avg. walking speed $v^*$; it is independent of the force $F_{cart}$ that must be applied to the cart. Since the avg. walking speed $v^*$ determines the avg. step length $s^*$, thus fixing the walking gait, the force $F_{cart}$ entirely determines the radius $r^*$. Combining (23) and (28) gives the external force:

$$F_{ext} = \sqrt{\left\langle \dot{W}_{gait}\left(v^*, s^*\right) \right\rangle / \varepsilon_{st}}. \tag{29}$$

Thus, given the walking gait, if the mechanical advantage is chosen according to equation C in (28), then the external force the subject must generate is constant.

*5.4 Predicted Movement Patterns*

We illustrate the maximum net energy efficiency walking gait and mechanical advantage derived in Sec. 5.3 using the case in which a subject selects a walking gait with the typical adult avg. walking speed $v^\circ$. We first use the parameter values in (17); we estimate:

$$\begin{aligned} s^* &\approx 0.67\,m, \\ r^*\left(F_{cart}\right)/R &\approx \left[130\,N\right]/F_{cart}, \\ F_{ext} &\approx 130\,N, \\ v_{cart}\left(F_{cart}\right) &\approx \left[170\,N \cdot m \cdot s^{-1}\right]/F_{cart}. \end{aligned} \tag{30}$$

For the cart pulled by Atzler & Herbst's subject requiring forces $F_{cart}$ of 100 N, 110 N, 130 N, and 160 N, the subject pulls the cart by generating an external force $F_{ext}$ of 130 N with the carts travelling at speeds $v_{cart}$ of 1.7 m·s⁻¹, 1.5 m·s⁻¹, 1.3 m·s⁻¹, and 1.1 m·s⁻¹, respectively. We can estimate the maximum external force the subject can generate as sustained exertion for this walking gait using the model maximum external force model developed in [13]; this gives:

$$F_{ext} \leq \left[370\,N\right] - mgs^*/2L \approx 230\,N. \tag{31}$$

Thus the external force $F_{ext}$ that the subject must generate to move the cart estimated in (30) is well below the maximum external force that we estimate for the subject in (31) given the walking gait the subject selects.

We next use the parameter values in (14) to provide an estimate on the how much the first model deviates from the corrected metabolic energy model for external forces $F_{ext}$ on the order of those observed in Atzler & Herbst; we estimate:

$$\begin{aligned} s^* &\approx 0.78\,m, \\ r^*\left(F_{cart}\right)/R &\approx \left[150\,N\right]/F_{cart}, \\ F_{ext} &\approx 150\,N, \\ v_{cart}\left(F_{cart}\right) &\approx \left[200\,N \cdot m \cdot s^{-1}\right]/F_{cart}. \end{aligned} \tag{32}$$

For the cart pulled by Atzler & Herbst's subject, the subject pulls the cart by generating an external force $F_{ext}$ of 150 N with the carts travelling at speeds $v_{cart}$ of 2.0 m·s⁻¹, 1.8 m·s⁻¹, 1.5 m·s⁻¹, and 1.3 m · s⁻¹, respectively. The estimated maximum external force the subject can generate as sustained exertion is:

$$F_{ext} \leq \left[370\,N\right] - mgs^* / 2L \approx 210\,N. \tag{33}$$

*5.5 Discussion*

An alternative approach to avoiding the very slow movement that happens when the subject simply tries to maximize the net energy efficiency (Sec. 5.2) than by fixing the avg. walking speed $v$ (Sec. 5.3) would be to fix the avg. speed $v_{cart}$ of the cart, effectively fixing the time the task takes to complete. The solution of this problem would follow an approach analogous to the one in Sec. 5.3. Fixing the avg. walking speed gives combinations of walking gait and mechanical advantage that fixes the external force the subject must generate though at the cost of possibly taking a long time to complete the task, while fixing the avg. speed $v_{cart}$ of the cart fixes the time taken to complete the task though at the cost of possibly using more onerous walking gaits and generating larger external forces.

The idealized mechanical device of massless and frictionless spools and cords is intended to provide clarity as to how mechanical advantage relates the task being performed to the external force that must be generated and the metabolic energies expended generating muscle forces and mechanical energy. We can replace the idealized mechanical device that we have used with a somewhat more practical, but still idealized one, one consisting of two spools of fixed radii each attached to a gear whose radius can be varied and with the two gears attached by a chain where the component parts are still massless and frictionless. This modified device begins to resemble the system of gears on bicycle and we see that the mechanical advantage becomes expressed in terms of the gear ratio.